\documentclass[preprintnumbers,amsmath,amssymb,aps,prl]{revtex4}
\usepackage{amsmath}
\usepackage[mathcal]{eucal}
\usepackage{latexsym}
\usepackage{amssymb}
\usepackage{color}
\newcommand*{\eps}{{\rlap{\lower2ex\hbox{$\,\,\tilde{}$}}{\epsilon_{ijk}}}}
\newcommand*{\EPS}{{\rlap{\lower2ex\hbox{$\,\,\tilde{}$}}{\epsilon_{i'j'k'}}}}
\newcommand*{\lmq}{{\rlap{\lower2ex\hbox{$\,\,\tilde{}$}}{\epsilon_{lmq}}}}
\newcommand*{\jmq}{{\rlap{\lower2ex\hbox{$\,\,\tilde{}$}}{\epsilon_{jmq}}}}
\newcommand*{\jql}{{\rlap{\lower2ex\hbox{$\,\,\tilde{}$}}{\epsilon_{jql}}}}
\newcommand*{\jlm}{{\rlap{\lower2ex\hbox{$\,\,\tilde{}$}}{\epsilon_{jlm}}}}
\newcommand*{\imq}{{\rlap{\lower2ex\hbox{$\,\,\tilde{}$}}{\epsilon_{imq}}}}
\newcommand*{\iql}{{\rlap{\lower2ex\hbox{$\,\,\tilde{}$}}{\epsilon_{iql}}}}
\newcommand*{\ilm}{{\rlap{\lower2ex\hbox{$\,\,\tilde{}$}}{\epsilon_{ilm}}}}
\newcommand*{\lmn}{{\rlap{\lower2ex\hbox{$\,\,\tilde{}$}}{\epsilon_{lmn}}}}
\newcommand*{\abc}{{\rlap{\lower2ex\hbox{$\,\,\tilde{}$}}{\epsilon_{abc}}}}
\newcommand*{\N}{{\rlap{\lower2ex\hbox{$\,\,\tilde{}$}}{N}}}
\newcommand{\tN}{{\rlap{\lower2ex\hbox{$\,\,\tilde{}$}}{N}}}
\newcommand*{\tM}{{\rlap{\lower2ex\hbox{$\,\,\tilde{}$}}{M}}}
\newcommand*{\imn}{{\rlap{\lower2ex\hbox{$\,\,\tilde{}$}}{\epsilon_{imn}}}}

\begin{document}
\title{Translational affine coherent states as solutions to the Wheeler-DeWitt equation}

\author{Eyo Eyo Ita III}\email{ita@usna.edu}
\address{Physics Department, US Naval Academy. Annapolis, Maryland}
\author{Chou Ching--Yi}\email{ccmaple014@gmail.com}
\address{National Center for Theoretical Sciences (South), National Cheng Kung University, Tainan, Taiwan}
\input amssym.def
\input amssym.tex

\bigskip

\begin{abstract}
The Quantum Wheeler-DeWitt operator can be derived from an affine
commutation relation via the affine group representation formalism for gravity, 
wherein a family of gauge-diffeomorphism invariant affine coherent states
are constructed from a fiducial state.  In this article, the role of the
fiducial state is played by a regularized Gaussian peaked on densitized triad
configurations corresponding to 3-metrics of constant spatial scalar 
curvature.  The affine group manifold consists of points in the upper 
half plane, wherein each point is labeled by two local gravitational 
degrees of freedom from the Yamabe construction. From this viewpoint, 
here we show that the translational subgroup of affine coherent states 
constitute a set of exact solutions of the Wheeler-DeWitt equation.
The affine translational parameter $b$ admits a physical interpretation 
analogous to a continuous plane wave energy spectrum, where the curvature
constant $k$ plays the role of the energy.  This result shows that the 
affine translational subgroup generates transformations in the curvature 
constant $k$ from the Yamabe problem, while $k$ is inert under the kinematic
symmetries of gravity.
\end{abstract}

\maketitle

\section{1. Introduction}

In \cite{SOOITACHOU} it was shown that the local Hamiltonian constraint of four dimensional General Relativity (GR) of Lorentian
signature can be written as an affine commutation relation
\begin{eqnarray}
\label{DELTA}
\widehat{H}(x)\vert\psi\rangle=\bigl([i\widehat{Q},\widehat{V}(x)]-\lambda\widehat{V}(x)\bigr)\vert\psi\rangle,\,\,\lambda=\frac{\hbar
G\Lambda}{2},\,\,G=\frac{8\pi G_{Newton}}{c^{3}}
\end{eqnarray}
\noindent wherein $\Lambda$ is the cosmological constant. Here,
$\imath Q$ refers to the imaginary part of the Chern--Simons
functional of the Ashtekar self-dual connection $A^a_i=\Gamma^a_i+iK^a_i$, where $\Gamma^a_i$ is the triad-compatible spin connection.\footnote{Index
conventions are that symbols $a,b,c,\dots$ from the beginning of the
Latin alphabet are internal $SO(3)$ indices, while symbols $i,j,k$
from the middle are spatial indices in 3-space $\Sigma$. Both sets
of indices take values $1--3$.} In densitized triad-extrinsic curvature conjugate
variables $(\widetilde{E}^{i}_{a},K^{a}_{i})$, this is given by
\begin{eqnarray}
\label{DELTA1} \imath Q=
\imath\int_{\Sigma}d^3y\Bigl(\frac{1}{2}\Bigl(\widetilde{R}^i_a[\Gamma]K^a_i+K^a_i\widetilde{R}^i_a[\Gamma]\Bigr)-\frac{1}{3!}\widetilde{\epsilon}^{ijk}\epsilon_{abc}K^a_iK^b_jK^c_k\Bigr)
\end{eqnarray}
\noindent and $V(x)$ is the local volume element of 3-space $\Sigma$,
given by
\begin{eqnarray}
\label{VOLUME}
V(x)=|\frac{1}{6}\widetilde{\epsilon}^{ijk}\epsilon_{abc}e^a_i(x)e^b_j(x)e^c_k(x)|
=\sqrt{\frac{1}{3!}\eps\epsilon^{abc}\widetilde{E}^{i}_{a}(x)\widetilde{E}^{j}_{b}(x)\widetilde{E}^{k}_{c}(x)}
\end{eqnarray}
\noindent The magnetic field of the spin connection $\Gamma^{a}_{i}$
compatible with spatial triads $e^{a}_{i}$ is given by
\begin{eqnarray}
\label{CURVATURE}
\widetilde{R}^i_a[\Gamma]=\widetilde{\epsilon}^{ijk}\partial_j\Gamma^a_i+\frac{1}{2}\widetilde{\epsilon}^{ijk}\epsilon_{abc}\Gamma^b_j\Gamma^c_k,
\end{eqnarray}
\noindent and $\widetilde{R}^i_a[\Gamma]e^a_i=\frac{eR}{2}$,
wherein $R=h^{ij}R_{ij}$ is the $3$-D Ricci scalar curvature of the
$3$-metric $h_{ij}=e^a_ie^a_j$.
\par \indent One of the main results of
\cite{SOOITACHOU} is the proposition of a physical Hilbert space
$\textbf{H}_{Phys}$ for four dimensional gravity of Lorentzian
signature according to the following prescription.  That is, if
there exists a gauge-diffeomorphism invariant fiducial state
$\vert\eta\rangle\equiv\vert{0},0\rangle$ forming a unitary
irreducible representation of the affine algebra (\ref{DELTA}), then
it should constitute a solution to the Hamiltonian constraint.  It follows 
from group theoretical considerations that if the fiducial state is a solution of the affine commutation
relation $\widehat{H}(x)\vert\eta\rangle=0$, then there exists a
family of gauge-diffeomorphism invariant affine coherent states $\vert{a},b\rangle=U(a,b)\vert\eta\rangle$, such that
$\widehat{H}(x)\vert{a},b\rangle=0$, wherein
\begin{eqnarray}
\label{ELEMENT}
U(a,b)=e^{ia\widehat{Q}}e^{ib\widehat{V}}
\end{eqnarray}
\noindent are unitary affine group elements formed from
exponentiation of the self-adjoint operators (\ref{DELTA1}) and
(\ref{VOLUME}) and we have defined global volume
$V=\int_{\Sigma}d^3xV(x)$. Affine coherent states
$\vert{a},b\rangle$ form an overcomplete basis for the physical
Hilbert space $\textbf{H}_{Phys}$, and to each point $(a,b)$ in the
upper half plane with $a>0$, which constitutes the group manifold,
there corresponds a single state. The fiducial state
$\vert{0},0\rangle\equiv\vert\eta\rangle$ acts as a seed that
generates the whole Hilbert space.
\par \indent While this is a
mathematical result of the representation theory of the affine
group, it must be shown that such a state $\vert\eta\rangle$ is indeed 
annihilated by the Wheeler--DeWitt constraint operator in order for
this representation theory to be of direct relevance to GR.  An explicit
example of a state which identically satisfies the Wheeler--DeWitt
equation $\widehat{H}(x)\vert\psi\rangle=0$ in this sense was found in
\cite{SOOITACHOU1}. The solution $\vert\psi\rangle$ consists of
states with support on densitized triads
$\widetilde{E}^i_a(x)=(\widetilde{E}^i_a(x))_R~\forall{x}\in\Sigma$
having a constant 3D spatial scalar curvature $R=6k$.  These states,
which are highly peaked about $(\widetilde{E}^i_a)_R$, can be
approximated by Gaussians in a certain regularization, which
approach the delta functional in the limit of removal of the
regulator.  The solutions of \cite{SOOITACHOU1}, which in general contain two
gravitational degrees of freedom per point of the spatial 3-manifold $\Sigma$, will serve for the purposes of the
present paper, as the wavefunctional in the densitized triad representation playing the
role of the fiducial state $\vert{0},0\rangle$.
\par \indent
 In the present paper we will show that the affine coherent states $\vert{a},b\rangle$ are as
 well annihilated by the Wheeler--DeWitt equation in densitized triad--extrinsic curvature
 variables $(\widetilde{E}^i_a,K^a_i)$ in a similar sense.  We will utilize the structures and
 definitions put in place in \cite{SOOITACHOU1}, including the Rigged Hilbert space construction
 applied to infinite dimensional functional spaces.  Our main result for this paper will be to
 explicitly demonstrate this feature in the functional Schrodinger representation for the
 translational elements $U(0,b)$ of the affine algebra, combined with their physical interpretation.
 The analogous computation for the dilational affine group element $U(a,0)$ will be relegated as a topic of future
 research.
 \par
\indent
This paper is organized as follows.  In Section 2 we recount some key results
from \cite{SOOITACHOU1}, specifically the regularized Wheeler--DeWitt equation in
self-adjoint form as descended from the affine Lie algebra, as well as the particular
functional Schrodinger representation to be used in this paper.  A key feature is the existence
of a regulator-induced singularity which resembles a cosmological constant.
In Section 3 we demonstrate the mechanism that generates the physical Hilbert space from a seed state.
We illustrate this at a formal level in Sections 3 and 4 using the properties of the affine algebra,
and rigorously in Section 3 for the translational part of the algebra.  We also provide a physical
interpretation for the meaning of affine group translations with respect to the Einstein equations.
A rigorous demonstration of the dilational transformation of the states has been relegated to a
future article, while in Section 4 we highlight some of the issues involved.  Section 5 is a brief
discussion section for our main results.

\section{2. The Hamiltonian constraint}

Substitution of (\ref{DELTA1}) and (\ref{VOLUME}) into (\ref{DELTA}) and reordering the terms into a particular self-adjoint form with kinetic operator schematically of the form $KeK$ leads to a regularized Wheeler--DeWitt equation given by \cite{SOOITACHOU1}
\begin{eqnarray}
\label{DELTA16} \widehat{H}_{\epsilon}(x)\psi[\widetilde{E}]
=\biggl[e\Bigl(\frac{R}{2}+\Lambda+\frac{9}{2}(\hbar{G}f_{\epsilon}(0))^2\Bigr)-\frac{1}{2}\widetilde{\epsilon}^{ijk}\epsilon_{abc}\widehat{K}^a_ie^b_k\widehat{K}^c_k\biggr]\psi[\widetilde{E}]
\end{eqnarray}
\noindent We will take
$\psi[\widetilde{E}]=\langle\widetilde{E}\vert\psi\rangle=\langle\widetilde{E}\vert\eta\rangle$
to be the fiducial state in the functional Schrodinger
representation diagonalized on the densitized triad
$\widetilde{E}^i_a$.  Note that there are regulator-dependent terms
$(f_{\epsilon}(0))^2$ which have been picked up as a result of the
reordering, which if the regulator were to be removed would blow up
as $(\delta^{(3)}(0))^2$.  We will follow the method of
\cite{SOOITACHOU1} in dealing with such terms. We will need the
equal-time commutation relations
\begin{eqnarray}
\label{COMMU}
[\widehat{\widetilde{E}^i_a}(x),\widehat{\widetilde{E}^j_b}(y)]=[\widehat{K}^a_i(x),\widehat{K}^b_j(y)]=0;~~
[\widehat{K}^a_i(x),\widehat{\widetilde{E}^j_b}(y)]=i(\hbar{G})\delta^a_b\delta^j_i\delta^{(3)}(x,y).
\end{eqnarray}
\noindent
There are an infinite number of possibilities for the functional Schrodinger representation.  We will start off by choosing one in which the triads act by multiplication and their conjugate momentum via functional differentiation
\begin{eqnarray}
\label{DELTA22}
\widehat{\widetilde{E}^i_a(x)}\psi[\widetilde{E}]=\widetilde{E}^i_a(x)\psi[\widetilde{E}];~~
\widehat{K}^a_i(x)\psi[\widetilde{E}]=i(\hbar{G})\frac{\delta\psi[\widetilde{E}]}{\delta\widetilde{E}^i_a(x)}.
\end{eqnarray}
\noindent
We will also choose a wavefunctional Ansatz of the form
\begin{eqnarray}
\label{ASSUME}
\psi[\widetilde{E}]=e^{S[\widetilde{E}]},
\end{eqnarray}
\noindent
for its facility of comparison with the semiclassical limit.  Then we consider the question, if $\psi$ were to be locally peaked about some configuration of the triads, what would that imply in terms of the Einstein equations.  From (\ref{COMMU}), one can readily read off the extrinsic curvature as
\begin{eqnarray}
\label{COMMU1}
\widehat{K}^a_i(x)\psi[\widetilde{E}]=i(\hbar{G})\Bigl(\frac{\delta{S}}{\delta\widetilde{E}^i_a(x)}\Bigr)\psi[\widetilde{E}].
\end{eqnarray}
\noindent
The result of acting with the triadic extrinsic curvature operator yields its Hamilton--Jacobi functional value times $\psi$.  Note that this value is a local function of $\widetilde{E}^i_a$ evaluated at the spatial point $x$.  Let us now invoke one of the properties of $\psi$ from \cite{SOOITACHOU1}, namely peakedness on a particular configuration $\widetilde{E}^i_a(x)=(\widetilde{E}^i_a(x))_R$ to be determined.  Then evaluated at $(e_{ia})_R$, we have
\begin{eqnarray}
\label{COMMU2}
i(\hbar{G})\Bigl(\frac{\delta{S}}{\delta\widetilde{E}^i_a(x)}\Bigr)_{e_R}=0.
\end{eqnarray}
\noindent
Equation (\ref{COMMU2}) is the condition that the semiclassical momentum is zero for precisely those configurations $e_R$ (namely $e^a_i(x)=(e^a_i(x))_R~\forall{x}\in\Sigma$ or alternatively, the same thing expressed in terms of densitized triads) for which $\psi$ is peaked.  Hence the configurations $(\widetilde{E}^i_a(x))_R$ constitute critical points of the functional $S$.\par
\indent
 Rewriting equation (\ref{DELTA16}) in the functional Schr\"odinger representation, we have a regularized Hamiltonian constraint given by
\begin{eqnarray}
\label{COMMU6}
\widehat{H}_{\epsilon}(x)\psi[\widetilde{E}]
=\biggl[e\Bigl(\frac{R}{2}+\Lambda+\frac{9}{2}(\hbar{G}f_{\epsilon}(0))^2\Bigr)+\frac{1}{2}(\hbar{G})^2\widetilde{\epsilon}^{ijk}\epsilon_{abc}
\Bigl(\frac{\delta}{\delta\widetilde{E}^i_a(x)}e^b_j(x)\frac{\delta}{\delta\widetilde{E}^k_c(x)}\Bigr)_{reg}\biggr]\psi[\widetilde{E}]=0,
\end{eqnarray}
\noindent
the subscript $reg$ denoting the regularization prescription of \cite{SOOITACHOU1} involving the regulating function $\tilde{f}_{\epsilon}(x,y)$ to handle the double-functional derivatives at the same spatial point $x$.  Hence we have that
$\hbox{lim}_{\epsilon\rightarrow{0}}\tilde{f}_{\epsilon}(x,y)=\delta^{(3)}(x,y)$.  Using (\ref{ASSUME}), we get the following expansion in terms of $S$
\begin{eqnarray}
\label{COMMU7}
\widehat{H}_{\epsilon}(x)\psi[\widetilde{E}]
=\biggl[e\Bigl(\frac{R}{2}+\Lambda+\frac{9}{2}(\hbar{G}f_{\epsilon}(0))^2\Bigr)\nonumber\\
+\frac{1}{2}(\hbar{G})^2\widetilde{\epsilon}^{ijk}\epsilon_{abc}
\Bigl[e^a_i\Bigl(\frac{\delta{S}}{\delta\widetilde{E}^j_b}\Bigr)\Bigl(\frac{\delta{S}}{\delta\widetilde{E}^j_b}\Bigr)
+\Bigl(\frac{\delta{e}^b_j}{\delta\widetilde{E}^i_a}\Bigr)_{reg}\Bigl(\frac{\delta{S}}{\delta\widetilde{E}^k_c}\Bigr)
+e^a_i\Bigl(\frac{\delta^2S}{\delta\widetilde{E}^j_b\delta\widetilde{E}^k_c}\Bigr)_{reg}\biggr]\psi[\widetilde{E}]=0.
\end{eqnarray}
\noindent
There are a few observations which can be made regarding (\ref{COMMU7}).  The first term of the second line, in combination with the first line excluding the $(f_{\epsilon}(0))^2$ term, are Hamilton--Jacobi equation type terms which one would normally use in constructing a semiclassical solution.  The remaining terms in the second line will each acquire a factor of $f_{\epsilon}(0)$ on account of the regularization prescription, terms which at the semiclassical level are typically ignored. What we will construct in this paper, analogously to \cite{SOOITACHOU1}, is an exact quantum solution to all orders.

\section{3. The affine group Hilbert space generation mechanism}

In \cite{SOOITACHOU1} solutions to (\ref{COMMU7}) were explicitly constructed, corresponding to manifolds of constant spatial scalar curvature $R=6k$.  These solutions would, in the affine formalism, correspond to some fiducial state $\vert\eta\rangle=\vert{0},0\rangle$, constituting the origin of the affine group manifold.  We will now demonstrate the mechanism whereby the fiducial state seeds the generation of the whole affine group Hilbert space, which was proven in \cite{SOOITACHOU}.  We will carry out the analysis for the translational part $U(0,b)$ of the affine group, reserving the dilational part $U(a,0)$ for a future article.  We will focus first on those configurations $e=e_R$, upon which the functional $S$ is peaked.  Evaluated on this particular configuration we can make use of the critical point feature (\ref{COMMU2}), wherein the semiclassical Hamilton--Jacobi momentum vanishes, and the only momentum squared terms which need to be considered are the Hessian terms in the second line of (\ref{COMMU7}).  But first, let us formally demonstrate the invariance of the Wheeler--DeWitt equation under affine group transformations.\par
\indent
The following group action ensues upon exponentiation of the affine Lie algebra
\cite{SOOITACHOU}
\begin{eqnarray}
\label{GROUPIT}
e^{ia\widehat{Q}}\widehat{V}e^{-ia\widehat{Q}}=e^{\lambda{a}}\widehat{V};~~
e^{ib\widehat{V}}\widehat{Q}e^{-ib\widehat{V}}=\widehat{Q}-\lambda{b}\widehat{V}.
\end{eqnarray}
\noindent
One sees that the dilator element has the following representation on $V$
\begin{eqnarray}
\label{DILATOR}
Q=-\frac{i\lambda}{2}\bigl(V\frac{d}{dV}+\frac{d}{dV}V\bigr)=-\frac{i\lambda}{2}-i\lambda\frac{d}{d\hbox{ln}V}.
\end{eqnarray}
\noindent
The idea will be, starting from the condition $\widehat{H}(x)\vert\eta\rangle=0$ for some $\eta$ satisfying the Wheeler--DeWitt equation,  then we would like to verify whether or not $U(a,b)\vert\eta\rangle$ also satisfies the same equation.  Hence, we will show that
\begin{eqnarray}
\label{IDEA}
{U}(a,b)\widehat{H}(x)\vert\eta\rangle
=U(a,b)U(a,b)^{\dagger}\widehat{H}(x)U(a,b)\vert\eta\rangle=0
\end{eqnarray}
\noindent
where $U(a,b)^{\dagger}U(a,b)=1$.  So the aim will be to show $U(a,b)^{\dagger}\widehat{H}(x)U(a,b)=\widehat{H}(x)$, namely that the Hamiltonian constraint is invariant under affine group transformations.  We will first reinforce this at the level of the algebra (\ref{DELTA}), and then subsequently at the level of the Wheeler--DeWitt equation itself (\ref{DELTA16}).  Without loss of generality, we will carry this out for each tranformation separately.\par
\indent

\subsection{3.1 Affine group translations}

Checking first the translational part of the affine algebra
$U(0,b)=e^{ib\widehat{V}}$, and starting from (\ref{DELTA}) we have
\begin{eqnarray}
\label{IDEA1}
\bigl([i\widehat{Q},\widehat{V}(x)]-\lambda\widehat{V}(x)\bigr)e^{ibV}\vert\eta\rangle
=e^{ibV}e^{-ibV}\bigl([i\widehat{Q},\widehat{V}(x)]-\lambda\widehat{V}(x)\bigr)e^{ibV}\vert\eta\rangle\nonumber\\
=e^{ibV}\bigl([ie^{-ib\widehat{V}}\widehat{Q}e^{ib\widehat{V}},\widehat{V}(x)]-\lambda\widehat{V}(x)\bigr)e^{ib\widehat{V}}\vert\eta\rangle\nonumber\\
=\bigl([i\widehat{Q}+i\lambda{b}\widehat{V},\widehat{V}(x)]-\lambda\widehat{V}(x)\bigr)\vert{0},b\rangle
=\bigl([i\widehat{Q},\widehat{V}(x)]-\lambda\widehat{V}(x)\bigr)\vert{0},b\rangle,
\end{eqnarray}
\noindent
yielding the invariance of $\widehat{H}(x)$ under translations at the level of the affine algebra.  We have used that $[\widehat{V}(x),\widehat{V}]=0$.  To examine the implictions of this property, as well as to check it, as regards GR, we will next perform the analogous calculation at the level of the Wheeler--DeWitt equation.  Starting from (\ref{DELTA16}), we have
\begin{eqnarray}
\label{IDEA2}
\biggl[e\Bigl(\frac{R}{2}+\Lambda+\frac{9}{2}(\hbar{G}f_{\epsilon}(0))^2\Bigr)-\frac{1}{2}\widetilde{\epsilon}^{ijk}\epsilon_{abc}\widehat{K}^a_ie^b_k\widehat{K}^c_k\biggr]e^{ibV}\vert\eta\rangle\nonumber\\
e^{ibV}e^{-ibV}\biggl[e\Bigl(\frac{R}{2}+\Lambda+\frac{9}{2}(\hbar{G}f_{\epsilon}(0))^2\Bigr)-\frac{1}{2}\widetilde{\epsilon}^{ijk}\epsilon_{abc}\widehat{K}^a_ie^b_k\widehat{K}^c_k\biggr]e^{ibV}\vert\eta\rangle\nonumber\\
=e^{ibV}\biggl[e\Bigl(\frac{R}{2}+\Lambda+\frac{9}{2}(\hbar{G}f_{\epsilon}(0))^2\Bigr)-\frac{1}{2}\widetilde{\epsilon}^{ijk}\epsilon_{abc}
(e^{-ib{V}}\widehat{K}^a_ie^{ib{V}})e^b_k(e^{-ib{V}}\widehat{K}^c_ke^{ibV})\biggr]\vert{0},b\rangle
\end{eqnarray}
\noindent
Let us now evaluate the kinetic term in the functional Schrodinger representation.  We have
\begin{eqnarray}
\label{IDEA3}
e^{-ib\widehat{V}}\widehat{K}^a_ie^{ib\widehat{V}}
=e^{-ibV}i(\hbar{G})\frac{\delta}{\delta\widetilde{E}^i_a(x)}e^{ibV}
=i(\hbar{G})\frac{\delta}{\delta\widetilde{E}^i_a(x)}+i(\hbar{G})\frac{\delta{(ibV)}}{\delta\widetilde{E}}\nonumber\\
=i(\hbar{G})\Bigl(\frac{\delta}{\delta\widetilde{E}^i_a(x)}+\frac{ib}{2}e^a_i(x)\Bigr),
\end{eqnarray}
\noindent
where we have used $\delta{V}/\delta\widetilde{E}^i_a=\frac{e^a_i}{2}$.  So the kinetic term of (\ref{IDEA2}) is given by
\begin{eqnarray}
\label{IDEA4}
\frac{1}{2}(\hbar{G})^2\widetilde{\epsilon}^{ijk}\epsilon_{abc}
\Bigl(\frac{\delta}{\delta\widetilde{E}^i_a(x)}+\frac{ib}{2}e^a_i(x)\Bigr)e^b_j(x)
\Bigl(\frac{\delta}{\delta\widetilde{E}^k_c(x)}+\frac{ib}{2}e^c_k(x)\Bigr)\psi[\widetilde{E}]\nonumber\\
=\frac{1}{2}(\hbar{G})^2\widetilde{\epsilon}^{ijk}\epsilon_{abc}\biggl[\Bigl(\frac{\delta}{\delta\widetilde{E}^i_a}e^b_j\frac{\delta}{\delta\widetilde{E}^k_c}\Bigr)_{reg}\psi[\widetilde{E}]
+\frac{ib}{2}\Bigl(\frac{\delta}{\delta\widetilde{E}^i_a}e^b_je^c_k\psi[\widetilde{E}]\Bigr)_{reg}
+\frac{ib}{2}e^a_ie^b_j\frac{\delta}{\delta\widetilde{E}^k_c}\psi[\widetilde{E}]
-\frac{b^2}{4}e^a_ie^b_je^c_k\psi[\widetilde{E}]\biggr]\nonumber\\
=\frac{1}{2}(\hbar{G})^2\widetilde{\epsilon}^{ijk}\epsilon_{abc}\biggl[\Bigl(\frac{\delta}{\delta\widetilde{E}^i_a}e^b_j\frac{\delta}{\delta\widetilde{E}^k_c}\Bigr)_{reg}\psi[\widetilde{E}]
+ibe^a_ie^b_j\frac{\delta}{\delta\widetilde{E}^k_c}\psi[\widetilde{E}]
+\frac{ib}{2}\Bigl(\frac{\delta(e^a_ie^b_j)}{\delta\widetilde{E}^k_c}\Bigr)_{reg}\psi[\widetilde{E}]
-\frac{b^2}{4}e^a_ie^b_je^c_k\psi[\widetilde{E}]\biggr].
\end{eqnarray}
\noindent
Using the relations
\begin{eqnarray}
\label{IDEA5}
\widetilde{\epsilon}^{ijk}\epsilon_{abc}e^b_je^c_k=2\widetilde{E}^k_c;~~
\widetilde{\epsilon}^{ijk}\epsilon_{abc}e^a_ie^b_je^c_k=6e;~~
\Bigl(\frac{\delta\widetilde{E}^k_c}{\delta\widetilde{E}^k_c}\Bigr)_{reg}=9\tilde{f}_{\epsilon}(0),
\end{eqnarray}
\noindent
then we have a regularized momentum squared term given by
\begin{eqnarray}
\label{IDEA6}
\widehat{C}_{\epsilon}(x)\psi[\widetilde{E}]
=\frac{1}{2}(\hbar{G})^2\widetilde{\epsilon}^{ijk}\epsilon_{abc}\Bigl(\frac{\delta}{\delta\widetilde{E}^i_a}e^b_j\frac{\delta}{\delta\widetilde{E}^k_c}\Bigr)_{reg}\psi[\widetilde{E}]
+ib(\hbar{G})^2\widetilde{E}^k_c\frac{\delta\psi[\widetilde{E}]}{\delta\widetilde{E}^k_c}
+e\Bigl(\frac{9ib(\hbar{G})^2f_{\epsilon}(0)}{2}-\frac{3(\hbar{G}b)^2}{4}\Bigr)\psi[\widetilde{E}],
\end{eqnarray}
\noindent
where we have used $\tilde{f}_{\epsilon}(0)=e(x)f_{\epsilon}(0)$.\par
\indent
Putting the result (\ref{IDEA6}) into (\ref{IDEA2}), we have the following Wheeler--DeWitt operator equation, which can be interpreted as the previous one plus the second and third terms of (\ref{IDEA6})
\begin{eqnarray}
\label{IDEA7}
\widehat{H}_{\epsilon}(x)\vert{0},b\rangle=
\biggl[e\biggl(\frac{R}{2}+\Lambda+\frac{9}{2}(\hbar{G}f_{\epsilon}(0))^2+\frac{9ib(\hbar{G})^2f_{\epsilon}(0)}{2}-\frac{3(\hbar{G}b)^2}{4}\biggr)
+b(\hbar{G})\widetilde{E}^k_c\widehat{K}^c_k
-\frac{1}{2}\widetilde{\epsilon}^{ijk}\epsilon_{abc}\widehat{K}^a_ie^b_j\widehat{K}^c_k\biggr]\vert{0},b\rangle=0.
\end{eqnarray}
\noindent We need to find a wavefunctional annihilated by
(\ref{IDEA7}), including cancelation of its regularization-dependent
terms.  Using the form
\begin{eqnarray}
\label{FORM}
\psi[\widetilde{E}]=e^{S[\widetilde{E}]}=\langle\widetilde{E}\vert{0},b\rangle,
\end{eqnarray}
\noindent
combined with the second line of (\ref{COMMU7}), we have a regularized Wheeler-DeWitt equation given by
\begin{eqnarray}
\label{IDEA8}
\widehat{H}_{\epsilon}(x)\psi[\widetilde{E}]=
\biggl[e\biggl(\frac{R}{2}+\Lambda+\frac{9}{2}(\hbar{G}f_{\epsilon}(0))^2+\frac{9ib(\hbar{G})^2f_{\epsilon}(0)}{2}-\frac{3(\hbar{G}b)^2}{4}\biggr)+ib(\hbar{G})^2\widetilde{E}^k_c\frac{\delta{S}}{\delta\widetilde{E}^k_c}\nonumber\\
+\frac{1}{2}(\hbar{G})^2\widetilde{\epsilon}^{ijk}\epsilon_{abc}
\Bigl[e^a_i\Bigl(\frac{\delta{S}}{\delta\widetilde{E}^j_b}\Bigr)\Bigl(\frac{\delta{S}}{\delta\widetilde{E}^j_b}\Bigr)
+\Bigl(\frac{\delta{e}^b_j}{\delta\widetilde{E}^i_a}\Bigr)_{reg}\Bigl(\frac{\delta{S}}{\delta\widetilde{E}^k_c}\Bigr)
+e^a_i\Bigl(\frac{\delta^2S}{\delta\widetilde{E}^j_b\delta\widetilde{E}^k_c}\Bigr)_{reg}\biggr]\psi[\widetilde{E}]=0.
\end{eqnarray}
\noindent In analogy with \cite{SOOITACHOU1} there are two cases
which we need to analyse, namely those configurations $e\neq{e}_R$
and $e=e_R$ for some $e_R$ upon which $S$ has one or more critical
points.  We will consider first the case $e=e_R$.

\subsection{3.2 The peaked configurations $e=e_R$}

Note, due to (\ref{COMMU2}), that for any $\epsilon>0$ the first and
second terms in the second line of (\ref{IDEA8}) will always be
zero, as well as the last term on the first line.  While the
vanishing of the first functional derivative of $S$ for $e=e_R$ is a
semiclassical effect at the Hamilton--Jacobi level, the second
functional derivative will be nonzero, which is a quantum statement.
So the only momentum squared term that will contribute for these
peaked configurations is the third term in the second line of
(\ref{IDEA8}).  This term, which requires a regulator, is given by
\begin{eqnarray}
\label{COMMU8}
\Bigl(\frac{\delta^2S}{\delta\widetilde{E}^i_a(x)\delta\widetilde{E}^j_b(x)}\Bigr)_{reg}
=\int_{\Sigma}d^3y\tilde{f}_{\epsilon}(x,y)\Bigl(\frac{\delta^2S}{\delta\widetilde{E}^i_a(x)\delta\widetilde{E}^j_b(y)}\Bigr)_{e_R}=\tilde{f}_{\epsilon}(0)M^{ab}_{ij},
\end{eqnarray}
\noindent where $M^{ab}_{ij}$ is the Hessian matrix of the
functional $S$.
\par \indent Upon naive removal of the regulator, the
Hamiltonian constraint (\ref{IDEA8}) would blow up for $e=e_R$. To
avoid this, all regulator-dependent terms must somehow cancel out so
that the constraint is identically satisfied, free of infinities.
The only way possible to achieve this seems to be that the
functional $S$ must acquire some dependence upon the regulator.  So
let us further refine the wavefunctional Ansatz from $\psi=e^S$ to
\begin{eqnarray}
\label{COMMU9}
\psi_{\epsilon}[\widetilde{E}]=\hbox{exp}\Bigl[\Bigl(\alpha{f}_{\epsilon}(0)+\beta+\frac{\gamma}{f_{\epsilon}(0)}\Bigr)S\Bigr]
\end{eqnarray}
\noindent where $\alpha$, $\beta$ and $\gamma$ are numerical
constants to be fixed by the requirement that the constraint be
identically satisfied.  Performing the replacement of (\ref{COMMU9})
in (\ref{IDEA8}), recalling that we are in the $e=e_R$ case, then
the surviving terms of the regularized Hamiltonian constraint are
\begin{eqnarray}
\label{COMMU10}
\widehat{H}_{\epsilon}(x)\psi_{\epsilon}[\widetilde{E}]=e\Bigl[\frac{R}{2}+\Lambda
+\frac{1}{2}(\hbar{G}f_{\epsilon}(0))^2\bigl(9+\alpha M\bigr)
 +(\hbar{G})^2\Bigl(f_{\epsilon}(0)\bigl(\beta{M}+\frac{9ib}{2}\bigr)+\gamma{M}-\frac{3b^2}{4}\Bigr)\Bigr]_{e_R}\psi_{\epsilon}[\widetilde{E}]=0
\end{eqnarray}
\noindent
where we have defined
\begin{eqnarray}
\label{COMMU111}
M=\widetilde{\epsilon}^{ijk}\epsilon_{abc}e^a_iM^{bc}_{jk}
\end{eqnarray}
\noindent
evaluated at $e_R$.  To avoid the blowing up of (\ref{COMMU10}) as $\epsilon\rightarrow{0}$ we must choose
\begin{eqnarray}
\label{NOTE}
\alpha=-\frac{9}{M};~~\beta=-\frac{9ib}{2M},
\end{eqnarray}
which eliminates the $(f_{\epsilon}(0))^2$ and $f_{\epsilon}(0)$ terms, $\forall\epsilon$, leaving us with the following condition which must be satisfied
\begin{eqnarray}
\label{NOTE1}
\Bigl(\frac{R}{2}+\Lambda-\frac{3(\hbar{G}b)^2}{4}+(\hbar
G)^{2}\gamma
M\Bigr)\psi_{\epsilon}[\widetilde{E}]\biggl\vert_{e=e_R}=0.
\end{eqnarray}
\noindent It is shown in \cite{SOOITACHOU1} that a choice of
$\gamma$ is tantamount to a choice of curvature. But the terms which
is quadratic in $b$ allow the freedom to change that curvature via
unitary transformation using the affine group translational element $U(0,b)=e^{ibV}$.  So
without loss of generality, we will take $\gamma=0$.  So the peaked
configurations identically satisfy the regularized Wheeler--DeWitt
equation for $all$ $\epsilon>0$, and these configurations have
support on triads $e_R$ such that
\begin{eqnarray}
\label{SUPPORT}
\frac{R}{2}+\Lambda-\frac{3(\hbar{G}b)^2}{4}=0\longrightarrow{R=6k},
\end{eqnarray}
\noindent
for curvature constant $k$.  This implies that
\begin{eqnarray}
\label{SUPPORT1}
b=\frac{2}{\hbar{G}}\sqrt{\frac{\Lambda}{3}+k}\longrightarrow{U}(0,b)=e^{i(2/\hbar{G})\sqrt{\Lambda/3+k}V},
\end{eqnarray}
\noindent which provides the meaning of the affine group
translational parameter $b$.  It is directly related to the
curvature constant $k$ appearing in the Yamabe problem.  So the
state $\vert{0},b\rangle=e^{ibV}\vert{0},0\rangle$ is annihilated by
$\widehat{H}(x)$ provided that $\widehat{H}(x)$ annihilates the
wavefunctional
\begin{eqnarray}
\label{NOTE3} \psi_{\epsilon}[\widetilde{E}]
=\hbox{exp}\Bigl[-\frac{1}{M}\Bigl(9f_{\epsilon}(0)+\frac{9ib}{2}\Bigr)\Bigr]
=e^{-(9ib/2M)S}\hbox{exp}\Bigl[\frac{-9f_{\epsilon}(0)}{M}S\Bigr].
\end{eqnarray}
\noindent We can make the identification
$\langle\widetilde{E}\vert{0},0\rangle=e^{-9\frac{f_{\epsilon}(0)}{M}S}$
in (\ref{NOTE3}), which satisfies the original Hamiltonian
constraint
$\widehat{H}_{\epsilon}(x)\vert{0},0\rangle_{\epsilon}=0$.  Hence
the regularized constraint $\widehat{H}_{\epsilon}(x)$ for the
translated case must be satisfied by
\begin{eqnarray}
\label{MUSTBE}
\langle\widetilde{E}\vert{0},b\rangle_{\epsilon}=e^{ibV}e^{-(9ib/2M)S}e^{-(9f_{\epsilon}(0)/{M})S}.
\end{eqnarray}
\noindent
We would have a confirmation of the affine translation mechanics via the desired form of $\vert{0},b\rangle=e^{ibV}\vert{0},0\rangle$ in (\ref{MUSTBE}) if not for the factor $e^{-(9ib/2M)S}$.  This factor was picked up in order to cancel a certain infinity, and is a result of having to introduce the regulator.  We must now remove the regulator, and recall from \cite{SOOITACHOU1} that in the case of the Gaussian, we have
\begin{eqnarray}
\label{MUSTBE1}
e^{-(9f_{\epsilon}(0)/{M})S}\longrightarrow\delta(\widetilde{E}-\widetilde{E}_R)\equiv\prod_{x\in\Sigma}\prod_{i,a}\delta(\widetilde{E}^i_a(x)-(\widetilde{E}^i_a(x))_R).
\end{eqnarray}
\noindent The delta functional has support on configurations
$e=e_R$, vanishing for all $e\neq{e}_R$ for $\epsilon=0$. But along
with the Gaussian in this regularization prescription, as per
(\ref{MUSTBE}) also comes the factor $e^{-(9ib/2M)S}$, which has no
direct bearing on the affine algebra representations.  The existence
of regularization dependence would be an undesirable effect within
the affine algebra.  However, it must be the case that
\begin{eqnarray}
\hbox{lim}_{\epsilon\rightarrow{0}}e^{-(9ib/2M)S}e^{-(9f_{\epsilon}(0)/{M})S}=\hbox{lim}_{\epsilon\rightarrow{0}}e^{-(9f_{\epsilon}(0)/{M})S},
\end{eqnarray}
\noindent due to the delta functional confining of the support to
configurations $e=e_R$. So in the limit of removal of the regulator,
the unwanted $e^{-(9ib/2M)S}$ term becomes suppressed for
$e\neq{e_R}$ and is exactly equal to $1$ for $e=e_R$, since $S$ is a
quadratic functional $\widetilde{E}^i_a$ centered on
$(\widetilde{E}^i_a)_R$.  So we have upon removal of the regulator
that
\begin{eqnarray}
\label{MUSTBE2}
\langle\widetilde{E}\vert{0},b\rangle=e^{ibV}\delta(\widetilde{E}-\widetilde{E}_R)=\langle\widetilde{E}\vert{e}^{ib\widehat{V}}\vert{0},0\rangle
\end{eqnarray}
\noindent
exactly as dictated by the affine algebra.\par
\indent
\subsection{The nonpeaked configurations $e\neq{e}_R$}
Having shown that the $e=e_R$ configuration solves the
Wheeler--DeWitt equation $\forall\epsilon>0$ we must now consider
the $e\neq{e}_R$ configurations.   We will now utilize the
properties of the functional Schwartz space as defined in
\cite{SOOITACHOU1}, namely the set of functionals of rapid decrease.
For these configurations the quantum terms in the second line of
(\ref{IDEA8}) would blow up if the regulator were to be removed.
But since the form of the wavefunctional
$\psi=e^{S}\longrightarrow{e}^{-\alpha{f}_{\epsilon}S}$ has been
fixed by the $e=e_R$ case, it is clear that any possible
$f_{\epsilon}(0)$ terms in the Hamiltonian constraint will not blow
up any faster than $\psi$ goes to zero.  So for $e\neq{e}_R$ the
condition
$\widehat{H}_{\epsilon}(x)\psi_{\epsilon}[\widetilde{E}]=0$ will be
violated for all $\epsilon>0$, but the size of the violation can be
made arbitrarily small as $\epsilon\rightarrow{0}$, corresponding to
removal of the regulator.  It is precisely at $\epsilon=0$ that the
$e\neq{e}_R$ configurations are infinitely suppressed, resulting in
an exact solution to the constraint.

\subsection{3.3 Physical interpretation}

We have proven that $\vert{0},b\rangle\in{Ker}\widehat{H}(x)$ as a
consequence of $\widehat{H}(x)\vert{0},0\rangle$, and in the process
have determined the physical interpretation of the parameter $b$ via
(\ref{SUPPORT1}).  The requirement that the Hamiltonian constraint
be identically satisfied in (\ref{IDEA8}) at the critical point
$(\delta{S}/\delta\widetilde{E}^i_a)_{e=e_R}=0$ for arbitrary $b$
provides the following physical interpretation.  Recall from
\cite{SOOITACHOU1} that the constant curvature condition $R=6k$ is a
gauge-diffeomorphism invariant statement.  So two curvatures $k$ and
$k+dk$, however, close together, cannot be related by a
diffeomorphism or gauge transformation connected to the identity. In
this sense, the solution to the Yamabe problem partitions 3-metrics
$h_{ij}$ into distinct equivalence classes labeled by the curvature
$k$.  So for any given curvature $k$, one is confined to that value
of $k$ and the accompanying two gravitational degrees of freedom per
point associated with the Yamabe problem for curvature $6k$,
provided that one is restricted to the $SO(3)$ and diffeomorphism
transformations.  The label $k$ then, by definition, refers to all
solutions within the same gauge and diffeomorphism equivalence
class.

\par \indent The condition (\ref{IDEA8}) being satisfied
corresponds to a critical point of $S$, which means that affine
translation with respect to the parameter $b$ results in a change of
the curvature $k$.  This transformation $U(0,b)$ as mentioned above
is neither a $SO(3)$ gauge transformation nor a spatial
diffeomorphism.  Rather, $U(0,b)$ is an affine group translation,
which is a transformation induced by the fact that the Hamiltonian
constraint, affine algebra (\ref{DELTA}), has an exact group
theoretical solution.  So translation in the $b$ direction in the
affine group manifold coordinatized by points in the upper half
plane $(a,b)$ corresponds to a transformation of the curvature $k$.
Whether the affine transformation should be considered an additional
gauge symmetry of GR which generates transformations of the
curvature constant $k$, or whether distinct curvatures correspond to
distinct physical configurations will be an interesting topic for
future discussion. Nevertheless, it seems to suggest that this could
be a possible mechanism by which one may distinguish transformations
generated by the Hamiltonian constraint as being physical rather
than gauge.

\par \indent A nice analogy happens in spontaneous
symmetry breaking of $\lambda\phi^4$ theory at the level prior to
the breaking of the symmetry.  The symmetry signifies a degenerate
vacuum of the theory, where the field $\phi$ rolls in a the valley
of a $\phi^4$ potential with mass squared term of the right sign. In
the same way that the symmetry continuously connects points on the
vacuum manifold, so do the affine translations continuously connect
configurations of constant curvature.

\par \indent Finally, the
identification of the solution
\begin{eqnarray}
\label{IDENTIFICATION}
\langle\widetilde{E}\vert{U}(0,b)|\psi\rangle=e^{i(2/\hbar{G})\sqrt{\Lambda/3+k}V}\delta(\widetilde{E}-\widetilde{E}_R)
\end{eqnarray}
\noindent corresponds to unitary affine group translations via the
aforementioned mechanism for $\frac{\Lambda}{3}+k>0$.  This is
analogous to the classical evolution of a free point particle
$e^{(i/\hbar)E_{k}{t}}$ with continuous energy eigenvalue labeled as
$E_{k}$, playing a role analogous to the curvature label $k$.
Moreover, we can have superpositions of states with different values
of $E_{k}$ and consider analogies to problems in ordinary one
dimensional quantum mechanics (here, the volume $V$ is playing the
role of the time $t$), therefore, (\ref{IDENTIFICATION}) can be
re-formulated as
\begin{eqnarray}
\label{IDENTIFICATIONa}
\langle\widetilde{E}\vert{U}(0,b)|\psi\rangle=e^{\frac{\imath}{\hbar}[\frac{2}{G}\sqrt{\frac{\Lambda}{3}+k}]t}\delta(\widetilde{E}-\widetilde{E}_R)
=e^{\imath bV}\delta(\widetilde{E}-\widetilde{E}_{R})
\end{eqnarray}
which shows the translational parameter $b$ depends on constant
curvature $k$ and for every particular constant curvature $k$, it
gives the energy
\begin{eqnarray}
E_{k}=b\hbar=\frac{2}{G}\sqrt{\frac{\Lambda}{3}+k}
\end{eqnarray}
 But for $k<-\frac{\Lambda}{3}$, the wavefunctional goes as
\begin{eqnarray}
\label{IDENTIFICATION}
\langle\widetilde{E}\vert{U}(0,b)|\psi\rangle\sim{e}^{-(2/\hbar{G})\sqrt{\vert\Lambda/3+k\vert}V}\delta(\widetilde{E}-\widetilde{E}_R),
\end{eqnarray}
\noindent
which is some type of tunneling solution.  A consequence of this is that quantum-mechanically, due to the affine group, the expansion of spacetimes with highly negative curvature must be suppressed, in relation to universes with small volume.

\section{4. Affine group dilations}

We have carried out the main goal of this paper, namely to establish
a physical interpretation for the translational part of the affine
Lie algebra with respect to GR solutions.  Displacement in $b$ of
the group manifold corresponds to a change in the constant curvature
$k$.  The dilational part of the affine group will be somewhat more
involved to establish in the functional Schrodinger representation.
The corresponding generator (\ref{DELTA1}) in this representation is
given by
\begin{eqnarray}
\label{HARD}
i\widehat{Q}=i\int_{\Sigma}d^3y\biggl[\frac{1}{2}(\hbar{G})\Bigl(\widetilde{R}^i_a(y)\frac{\delta}{\delta\widetilde{E}^i_a(y)}
+\frac{\delta}{\delta\widetilde{E}^i_a(y)}\widetilde{R}^i_a(y)\Bigr)
-\frac{1}{3!}(\hbar{G})^3\widetilde{\epsilon}^{ijk}\epsilon^{abc}
\frac{\delta^3}{\delta\widetilde{E}^i_a(y)\delta\widetilde{E}^j_b(y)\delta\widetilde{E}^k_c(y)}\biggr].
\end{eqnarray}
\noindent
The facility of the translational case stems from the fact that the translational group element $U(0,b)=e^{ibV}$ acts by multiplication in the functional Schrodinger representation.  For dilations, the operator $\widehat{Q}$ acts by functional differentiation.  The Weyl-ordered version (\ref{HARD}) appears ghastly for numerous reasons.  First, it involves multiple functional derivatives acting at the same spatial point.  For a generic wavefunctional this could entail a regularization prescription of the form
\begin{eqnarray}
\label{HARD1}
\frac{i(\hbar{G})}{2}\int_{\Sigma}d^3y\widetilde{R}^i_a(y)\frac{\delta}{\delta\widetilde{E}^i_a(y)}
+\frac{i(\hbar{G})}{2}\int_{\Sigma}d^3x\int_{\Sigma}d^3y\tilde{f}_{\epsilon}(x,y)\frac{\delta\widetilde{R}^i_a(y)}{\delta\widetilde{E}^i_a(x)}\nonumber\\
-\frac{i}{3!}(\hbar{G})^3\int_{\Sigma}d^3x\int_{\Sigma}d^3y\int_{\Sigma}d^3z
\widetilde{\epsilon}^{ijk}\epsilon^{abc}
\tilde{f}_{\epsilon}(x,y)\tilde{f}_{\epsilon}(x,z)\frac{\delta^3}{\delta\widetilde{E}^i_a(x)\delta\widetilde{E}^j_b(y)\delta\widetilde{E}^k_c(z)}.
\end{eqnarray}
\noindent
The burden of accurate accounting of the regularization dependent terms of $\widehat{Q}$, is compounded infinity-fold by the fact that the operator must be exponentiated.  To this end it helps that $\widehat{Q}$ is self-adjoint.  Still it appears nontrivial, by any standard, to check that the Wheeler--DeWitt operator acting on a generic state is invariant under transformations generated by $\widehat{Q}$.  However, due to the results of \cite{SOOITACHOU}, when restricted to states forming a representation of the affine algebra, the action of (\ref{HARD}) to all orders, which corresponds to the limit of the action of the exponentiated version (\ref{HARD1}) when the regulator is removed, should by definition leave $\widehat{H}(x)$ in (\ref{DELTA}) invariant.  We will relegate an explicit demonstration of this proposition via the functional Schrodinger representation to a separate paper.  For the purposes of the present paper it will suffice to demonstrate this result formally using the properties of the affine algebra.\par
\indent
We would like to show that the Hamiltoinian constraint is invariant under affine group dilations, so that $\vert{a},0\rangle$ is a solution given that $\vert{0},0\rangle$ is a solution.  So we have, restricted to the space of physical states that
\begin{eqnarray}
\label{HARD2}
\widehat{H}(x)\vert{a},0\rangle=[i\widehat{Q},\widehat{V}(x)]\vert{a},0\rangle-\lambda\widehat{V}(x)\vert{a},0\rangle
=[i\widehat{Q},\widehat{V}(x)]e^{ia\widehat{Q}}\vert{0},0\rangle-\lambda\widehat{V}(x)e^{ia\widehat{Q}}\vert{0},0\rangle.
\end{eqnarray}
\noindent
Using the requirement that $[\widehat{Q},\widehat{Q}]=0$ for any operator irrespective of however ill-defined it may be, then (\ref{HARD}) can be rewritten as
\begin{eqnarray}
\label{HARD3}
\widehat{H}(x)\vert{a},0\rangle
=e^{ia\widehat{Q}}e^{-ia\widehat{Q}}[i\widehat{Q},\widehat{V}(x)]e^{ia\widehat{Q}}\vert{0},0\rangle
-{\lambda}e^{ia\widehat{Q}}e^{-ia\widehat{Q}}\widehat{V}(x)e^{ia\widehat{Q}}\vert{0},0\rangle.\nonumber\\
=e^{ia\widehat{Q}}[i\widehat{Q},e^{-ia\widehat{Q}}\widehat{V}(x)e^{ia\widehat{Q}}]\vert{0},0\rangle
-{\lambda}e^{ia\widehat{Q}}e^{-ia\widehat{Q}}\widehat{V}(x)e^{ia\widehat{Q}}\vert{0},0\rangle.
\end{eqnarray}
\noindent
It is at this point that we invoke the powerful results of the affine Lie algebra, namely that on states forming representations, we have the dilation
\begin{eqnarray}
\label{HARD4}
e^{-ia\widehat{Q}}\widehat{V}(x)e^{ia\widehat{Q}}=e^{-\lambda{a}}\widehat{V}(x)~\forall{x}\in\Sigma.
\end{eqnarray}
\noindent
This holds true irrespective of the form of the Lie algebra generators, and the proposition is that it must therefore hold true for (\ref{HARD}) restricted to elements of the physical Hilbert space $\textbf{H}_{Phys}$.  Importing (\ref{HARD4}) into (\ref{HARD3}), we have
\begin{eqnarray}
\label{HARD5}
H(x)|a,0\rangle=e^{ia\widehat{Q}}\Bigl[[i\widehat{Q},e^{-\lambda{a}}\widehat{V}(x)]\vert{0},0\rangle
-{\lambda}e^{-\lambda{a}}\widehat{V}(x)\vert{0},0\rangle\Bigr]=e^{-\lambda{a}}e^{ia\widehat{Q}}\widehat{H}(x)\vert{0,0}\rangle.
\end{eqnarray}
\noindent
So if $\widehat{H}(x)\vert{0},0\rangle=0$, then it follows that $\widehat{H}(x)\vert{a},0\rangle=0$ as well, namely that the Hamiltonian constraint is invariant under affine group dilations by $U(a,0)$.  Combined with the translational result, it follows from the representation space affine group Lie algebra (\ref{DELTA}) that the Hamiltonian constraint is invariant under affine group displacements $U(a,b)$.  Given that this is true at the group-theoretical level, we propose that this must as well be the case in the functional Schrodinger representation of the Wheeler--DeWitt equation.

\section{5. Discussion}
We have provided some concreteness to the meaning of affine algebra
with respect to gravity and solutions to Einstein equations.  We
have verified the result proven in \cite{SOOITACHOU} that affine
group forms an unitary, irreducible representation of the physical
Hilbert space.  Specifically, we have shown that Wheeler--DeWitt
operator $\widehat{H}(x)$ is invariant under affine group
transformations $U(a,b)$, formally at the level of affine algebra.
While we have relegated an explicit demonstration of this property
for the dilational group elements for future work, we have done so
rigorously for affine group translations.  This has provided a
physical interpretation for the meaning of the translational group
parameter $b$ as determined by the curvature constant $k$ for the
spatial manifold in question. Recall that a constant $k$ is inert
under $SO(3)$ gauge transformations and spatial diffeomorphisms.
Whether configurations corresponding to different values of $k$ are
physically equivalent or rather, physically distinct, will form an
interesting question for future research.  Finally, having
determined the interpretation of the constant $b$ as representing
something analogous to the energy eigenvalue of a free particle, it
remains to determine the corresponding interpretation for the
dilation parameter $a$.  We will relegate this determination as well
for future research.

\section{Acknowledgements}
This work has been supported in part by the
Office of Naval Research under Grant No. N-000-1414-WX-20789, and
in part by Perimeter Institute for Theoretical Physics (research at Perimeter
Institute is supported by the Government of Canada through Industry
Canada and by the Province of Ontario through the Ministry of Economic
Development and Innovation).  The research for this work also has
been supported by funds from the National Center for Theoretical
Sciences(South), Taiwan. Chou Ching-Yi would like to thank Professor
Chia-Chu Chen's support and encouragement.


\begin{thebibliography}{99}


\bibitem{SOOITACHOU} Chou C, Ita E and Soo C 2013 {\it Class.\ Quantum\ Grav.} {\bf 30}, 065013.

\bibitem{SOOITACHOU1} Ita E and Soo C 2014 Exact solutions to the Wheeler DeWitt equation and the Yamabe construction {\it
Preprint} gr-qc/1408.0710.
\end{thebibliography}
\end{document}